\documentclass[aps,prl,superscriptaddress,twocolumn,10pt,nofootinbib]{revtex4}
\usepackage[utf8]{inputenc}
\usepackage[english]{babel}
\usepackage{times}
\usepackage{color, soul}
\usepackage{graphicx}
\usepackage{import}
\usepackage{amsmath}
\usepackage{braket}
\usepackage{balance}

\usepackage{booktabs}

\AtBeginDocument{
\heavyrulewidth=.08em
\lightrulewidth=.05em
\cmidrulewidth=.03em
\belowrulesep=.65ex
\belowbottomsep=0pt
\aboverulesep=.4ex
\abovetopsep=0pt
\cmidrulesep=\doublerulesep{}
\cmidrulekern=.5em
\defaultaddspace=.5em
}

\begin{document}

\title{Multiplexed Pseudo-Deterministic Photon Source with Asymmetric Switching Elements}

\date{\today}

\author{Sebastian Brandhofer}
\email{sebastian.brandhofer@iti.uni-stuttgart.de}
\affiliation{Center for Integrated Quantum Science and Technology (IQST), University of Stuttgart, Stuttgart, Germany}
\author{Casey R. Myers}
\affiliation{Silicon Quantum Computing, Level 2, Newton Building, UNSW Sydney, 
Kensington, NSW, Australia}
\affiliation{School of Computing and Information Systems, Faculty of Engineering and Information Technology, The University of Melbourne, Melbourne VIC 3010, Australia}
\author{Simon J. Devitt}
\affiliation{Center for Quantum Software and Information, University of Technology Sydney, Ultimo, NSW, Australia}
\author{Ilia Polian}
\affiliation{Center for Integrated Quantum Science and Technology (IQST), University of Stuttgart, Stuttgart, Germany}

\begin{abstract}
The reliable, deterministic production of trustworthy high-quality single photons is a critical component of discrete variable, optical quantum technology.  For single-photon based fully error-corrected quantum computing systems, it is estimated that photon sources will be required to produce a reliable stream of photons at rates exceeding 1 GHz~\cite{4}.  Photon multiplexing, where low probability sources are combined with switching networks to route successful production events to an output, are a potential solution but requires extremely fast single photon switching with ultra-low loss rates.  In this paper we examine the specific properties of the switching elements and present a new design that exploits the general one-way properties of common switching elements such as thermal pads.  By introducing multiple switches to a basic, temporal multiplexing device, we are able to use slow switching elements in a multiplexed source being pumped at much faster rates.  We model this design under multiple error channels and show that anticipated performance is now limited by the intrinsic loss rate of the optical waveguides within integrated photonic chipsets.
\end{abstract}

\maketitle

Linear optical quantum computing was one of the earliest success stories in quantum computing research~\cite{17}.  Since the groundbreaking paper of Knill, Laflamme and Milburn in 2000, which demonstrated that linear optics and post-selected measurements~\cite{31} could be used to realise universal scalable quantum logic in discrete photonics, experimental demonstrations of small scale quantum algorithms have become routine around the world~\cite{2,23,14,13,7,10}.  Bulk optical experiments in the laboratory were replaced with more stable integrated optical photonic chips and algorithms consisting of up to 12 photonic qubits have been performed in a controlled manner~\cite{20}.  In recent years, photonic quantum computing systems have been adopted by several startups worldwide~\cite{36,35}, with the promise of building scalable single photon-based quantum computers~\cite{8}.  
\\
\\
However, single photon-based quantum computing platforms have always suffered from the reliable production of on-demand single photons.  The vast majority of experimental demonstrations of optical quantum computing use sources that are probabilistic, producing individual photons from the down-conversion of low amplitude coherent states from pulsed lasers (henceforth called weak laser pulse)~\cite{10}.  These non-deterministic sources are a major limiting factor for the demonstration of larger quantum algorithms in photonic platforms when millions or more photons are required in a temporally synchronous manner.  The push to build a scalable quantum computing system in optics is absolutely dependent on solving the source problem and constructing devices that can produce high-quality, identical, on-demand single photons. \\
\\
\\
In this paper, we detail a design for a new type of multiplexed source that exploits a specific property of integrated optical switches, namely their asymmetric switching properties.  Single photon switches, based on Mach-Zehnder interferometers and phase modulators, often have anti-symmetric phase profiles as a function of the control parameter---i.e. they effectively have different time scales associated with ``turning-on'' and ``turning-off'~\cite{16}. For example, using thermal pads to phase modulate a wave-guide in doped silicon can be switched ``on" quickly by adding thermal energy to the system, but removing that thermal energy to switch ``off" the modulator takes up to five times as long~\cite{3, 22}.  
\\
\\
Our new design---dubbed the racetrack source---takes this effect into consideration and allows for the construction of a multiplexed source where individual switches are only used once.  This allows for fast multiplexing using hardware switching that may be slow.  We perform detailed error modelling, and illustrate that a high-probability multiplexed source can be constructed.
\section{Background}
There are multiple approaches that have been proposed for building a device with the capability of producing high quality photons on-demand.  They can be broken down into two broad categories. 

\begin{itemize}
\item {\em Single photon emitters}: Matter-based quantum systems (such as atoms or quantum dots) that can be controlled and pulsed to emit single photons deterministically as the result of an energy transition~\cite{21,6,0,19,5}.
\item {\em Multiplexing}: Combining a large number of probabilistic sources with an active switching network to route successful photon generation events to the output of the device~\cite{12,25,9,11,18,32,1}.  
\end{itemize}

While single photon emitters may ultimately prove to be a more effective means of producing photons for quantum computing and communication application in the future, they currently come with several drawbacks.   The first is the ability to produce identical photons in artificial atoms such as quantum dots. In order for individual photons to be useful in optical quantum processing chipsets, they must interfere to high fidelity when mixed together on linear optical elements, such as beamsplitters.  For this to occur, individual photons must be close to 100\% indistinguishable in all degrees of freedom.  This is still difficult to achieve.  The physical fabrication of the solid-state emitter, its local environment and control inaccuracies all influence the ability to produce individual photons from  distinct  emitter sources that interfere with high accuracy on linear optical components.  Additionally, the integration of these components into linear optical quantum technology often diminishes the benefits of the optical platform as they are comparatively difficult and costly to fabricate at scale, can require complex infrastructure, such as dilution refrigeration cooling to operate, and are not always compatible with telecom frequencies when utilising long-range optical fibre systems as low loss quantum memories for scalable designs~\cite{34}. 
\\
\\
The current method of choice for photon production remains probabilistic downconversion processes such as in Spontaneous Parametric Down Conversion (SPDC) or Spontaneous Four Wave Mixing (SFWM) sources \cite{10} the latter of which has already been incorporated into integrated optical chipsets and experimentally utilised). These sources take a weak coherent laser source of frequency ($\omega$) and, through a weak optical nonlinearity, probabilistically generate a pair of lower frequency photons ($\omega_1+\omega_2 = \omega$).  The probability of success for these single photon sources is dictated by the power used in the pump laser.  The general output Gaussian state given a weak laser pulse input can be described by the sum over Fock states, 
\begin{equation}
    \sqrt{1-|\zeta|^2}\left( |vac\rangle_s|vac\rangle_i+\sum_{n=1}^\infty \zeta^N|n\rangle_s|n\rangle_i\right).
\end{equation}
The subscripts $s$ and $i$ refer to the signal and idler photons and $|\zeta|< 1$ parameterises the strength of the nonlinearity in the source and the power used in the pump, $|\zeta|=\tanh(cP)$, where $P$ is the power in mW and $c$ is the coupling constant of the optical nonlinearity in units of mW$^{-1/2}$~\cite{27}.  
\\
\\
The output of the source is consequently not a perfect single photon; we instead always get a superposition of the vacuum state and all Fock states.  As $|\zeta| < 1$, the higher-order Fock terms drop off exponentially.  Therefore, good single-photon sources are highly probabilistic.  This creates a  scaling problem.  For the production of two photons at the same time, 
two down-conversion sources must succeed at the same time, occurring with a probability of $p^2$.  If each source succeeds with a probability of 1/100, then two photons can only be produced with a probability of 1/10000.  This continues to decrease exponentially, such that producing $N$ photons at the same time for a quantum algorithm occurs with a probability of $p^N$.  
\\
\\
The solution to this problem is the concept of multiplexing: A large number of probabilistic sources are integrated with an active switching network so that when the device is triggered, on average, one of the probabilistic sources will be successful and the switching network can be configured, in real-time, such that the resultant photon can be routed to the output of the device.  However, the utility of this device to produce on-demand single photons is contingent on very accurate and fast single-photon switches.  These switches are difficult to fabricate reliably, and often a choice needs to be made between fast, lossy,  switches versus precise slow switches.  This is often due to the intrinsic asymmetry of the switches themselves.  While switches built, for example, from waveguide thermal pads~\cite{10} can be switched on very quickly (by dumping a large amount of thermal energy onto a chip in a very short time), they cannot be switched off quickly---essentially the system has to be allowed to cool through thermal dissipation. This limits either the speed or reliability of the switch itself and limits its use in a multiplexed source.  
\\
\\
Research has been ongoing to use different physical effects to induce the modulation needed to build a single photon switch---most notably electro-optical modulation in materials such as Lithium Niobate \cite{24}---but these systems are still in their technological infancy and are not straightforwardly compatible with the CMOS manufacturing.  This is important as CMOS compatibility and the ability to leverage global silicon manufacturing infrastructure for quantum technology is a big advantage for the utility of optics as a quantum hardware system.
\\
\\
In addition to intrinsic waveguide loss, inaccurate switching in these multiplexed sources is a large contributing factor to photon loss.  High loss rates or slow production of on-demand photons renders multiplexed photon sources effectively useless in the current generation of integrated optical systems, especially without the existence of high-quality and low-loss single-photon memories.  Photons need to be produced quickly, such that small amounts of fibre loops are sufficient to temporally synchronise photons as input to integrated optical computing chips and the pseudo-determinism of such a device has to be high to move into a regime where the probability of generating $N$ photons for an experiment scales at worst polynomially rather than exponentially in $N$. 
\\
\\
Our goal is to redesign a multiplexed switch specifically with this asymmetry in mind, using the CMOS-compatible thermal pad system to prototype a new design that operates each switch in the multiplexed system in a ‘one-way’ fashion.  As each switch is fast and accurate when switched ‘on’ from a previously ‘off’ position, we will construct a multiplexed switch where this is the only operation allowed for each switch in the device.  This leads to a source design that resembles a racetrack, with an array of switches coupling an inner waveguide loop to an outer loop that contains an array of probabilistic sources and heralding detectors.  The key insight is that each switch in the system can only be used ONCE, sidestepping the fundamental problem that makes switches in a multiplexed source either too unreliable or too slow.  
\\
\\
In this work, we present a novel multiplexed single photon source that addresses asymmetric configuration delays and corresponding error characteristics to produce high-quality photons with a high probability.

\section{Asymmetric Switch}

\begin{figure}[ht!]

    \centering
	
	\includegraphics[width=0.75\columnwidth]{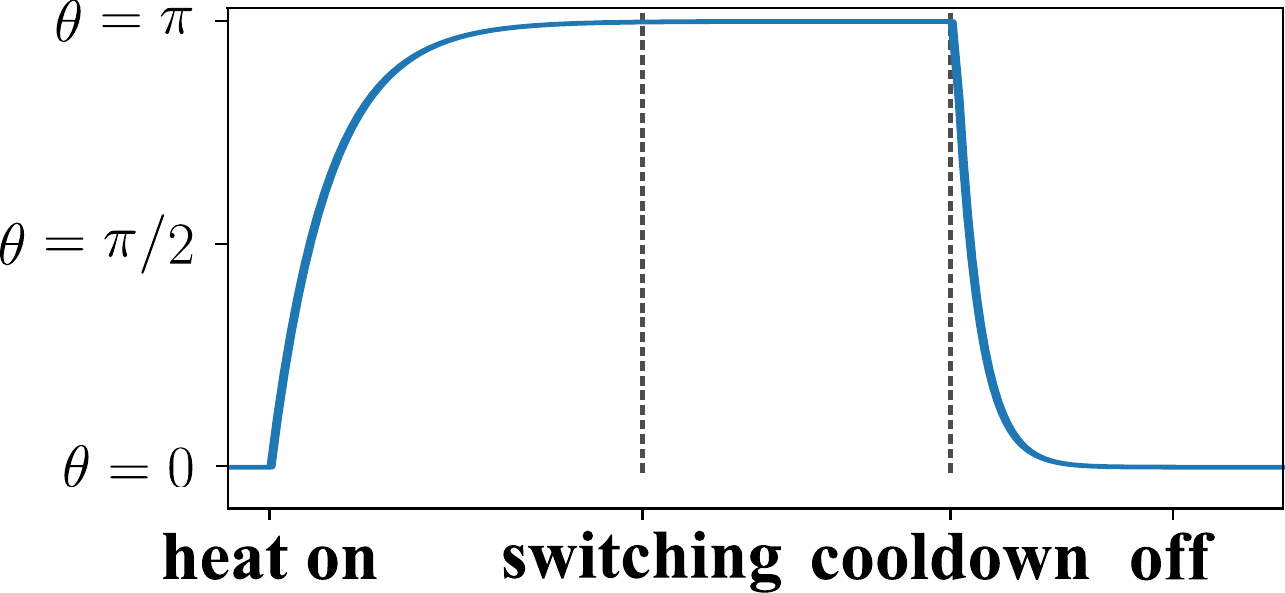}
	\caption{Schematic switching profile of switches based on thermo-optic effect.}
	\label{2}
\end{figure}

In our proposed single photon source each switch is configured at most twice.
This enables the exploitation of asymmetric configuration delays for switches with a specific switching profile as sketched in figure~\ref{2}.
For currently available switches based on thermo-optic effects, the switch configuration can often be changed once quickly e.g. by turning on a heat pad or by active cooling~\cite{16}.
However, as soon as the switch is configured once, reverting to the previous configuration requires a larger configuration delay.
According to the literature, the configuration delay $T_{sr}$ incurred by turning on a heat pad in thermo-optic switches may be approximately $3$ times smaller to $4$ times larger than $T_{sf}$, the configuration delay incurred by cooling down the heat pad~\cite{16}.
Regardless of $T_{sr}$ or $T_{sf}$ requiring more time, the polarity of the switches can be reversed to  exploit the asymmetric configuration delays for switching in the proposed single-photon source.
For the remainder of the paper, we assume that the time required to turn on a heat pad in a thermo-optic switch requires significantly less time than its cooldown.

\begin{figure}[ht!]
\centering
	\includegraphics[width=1.0\columnwidth]{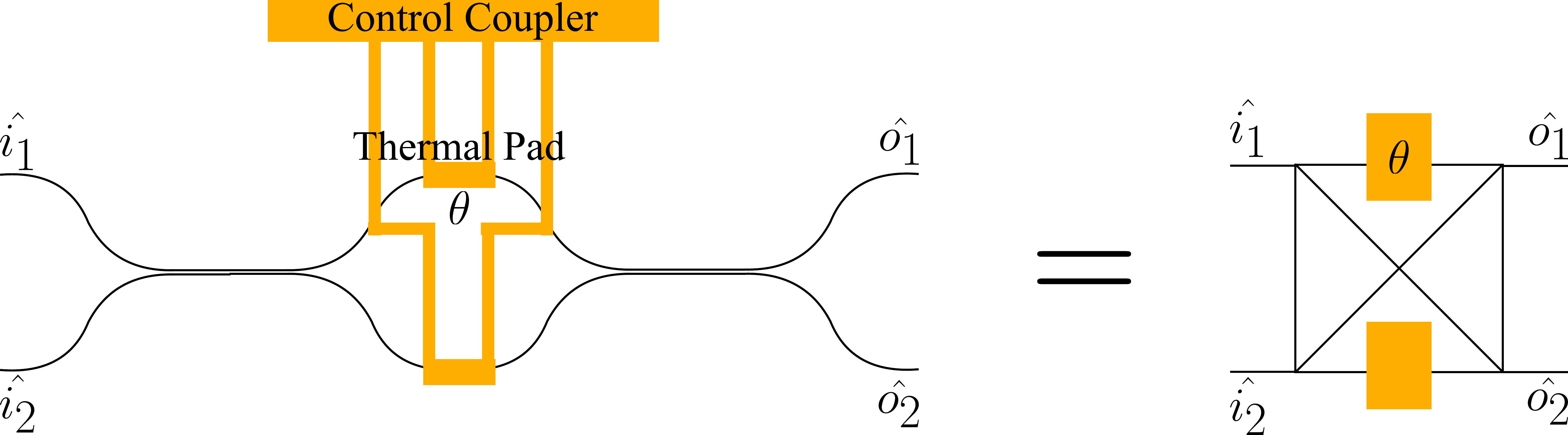}
	\caption{Switch based on thermo-optic effects with two thermal pads that can be activated individually.}
	\label{4}
\end{figure}
\subsection{Thermo-Optic Switches}
A single photon switch is a two-input, two-output device that allows for a single photon to be routed from either of the two inputs to either of the two outputs.  It has a generic structure shown in Fig. \ref{4}, which is drawn in the context of an integrated optics chipset. 
\\
\\
In Fig. \ref{4} we have three components, two directional couplers that work as single photon beamsplitters and a single device, known as a phase shifter, that acts to impart a relative phase on any photon that passes through the waveguides underneath.  To see how this switch works, we consider the transformation of one of the directional couplers.
\\
\\
If we label the optical modes of the input, $\hat{i_1}$ and $\hat{i_2}$ and the output of the first directional coupler as $\hat{t_1}$, $\hat{t_2}$, these modes will transform as,
\begin{equation}
\hat{i_1} = \frac{1}{\sqrt{2}}\left( \hat{t_1} + \hat{t_2}\right), \quad \hat{i_2} = \frac{1}{\sqrt{2}}\left(\hat{t_1}-\hat{t_2}\right)
\end{equation} 
These transformations also hold with the second directional coupler, instead now the inputs are modes $\hat{t_1}$ and $\hat{t_2}$ and we add two more outputs, $\hat{o_1}$ and $\hat{o_2}$.  
\\
\\
If we now combine two of these directional couplers together, it is easy to check that $\hat{i_1} \rightarrow \hat{o_1}$ and $\hat{i_2} \rightarrow \hat{o_2}$, i.e. a photon will {\em not be switched} when passed through the device.  However, if we now consider the phase shifter that is placed between the two directional couplers, that is designed to take $\hat{t_1} \rightarrow e^{i\theta} \hat{t_1}$, we find the input/output mapping of the switch is
\begin{equation}
\begin{aligned}
\hat{i_1} \rightarrow \frac{1}{2}\left( e^{i\theta} \hat{o_1} + e^{i\theta} \hat{o_2} + \hat{o_1} - \hat{o_2}\right) \\
\hat{i_2} \rightarrow \frac{1}{2}\left( e^{i\theta} \hat{o_1} + e^{i\theta} \hat{o_2} - \hat{o_1} + \hat{o_2}\right) \\
\label{5}
\end{aligned}
\end{equation}
Hence, for $\theta = 0$, Eq. \ref{5} takes $\hat{i_i} \rightarrow \hat{o_i}$, $i\in \{1,2\}$. If $\theta = \pi$, we get the transformations, $\hat{i_1} \rightarrow \hat{o_2}$ and $\hat{i_2} \rightarrow \hat{o_1}$ and any input photon will be switched to the opposite output. 
\\
\\
Thermo-optic switches can be used in figure~\ref{4}~that contain two thermal pads, each applicable to one of the connected waveguides separately. The induced phase shift $\theta$ can be manipulated by each heat pad individually. The initial switch configuration can be reverted in time $T_{sr}$ by activating the second heat pad.
However, changing the switch configuration a third time would then require the cooldown of one or both heat pads, leading to the delay $T_{sf}$.
The manufacturing of the proposed double-padded switches has been demonstrated in~\cite{16}. 
However, a characterisation of its exact switching profile and asymmetric error characteristics is pending.

\section{Multiplexed Single Photon Source}
\begin{figure*}[ht!]

	\centering
	\includegraphics[width=0.7\textwidth, draft=False]{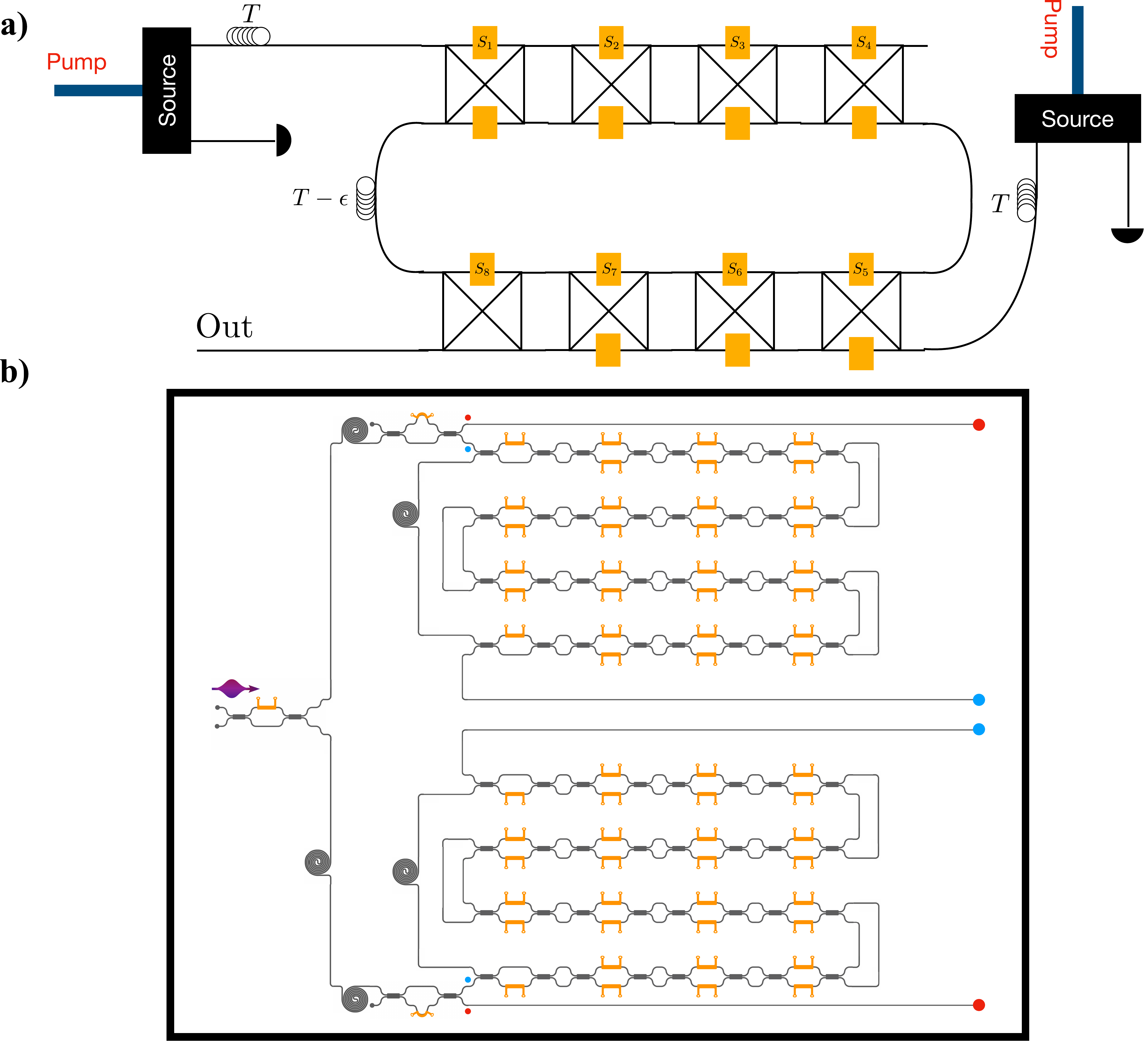}
	\caption{Developed photon source with two SPDC sources, seven double-padded switches that connect the sources to an inner delay loop and one single-padded switch $S_{8}$ that connects the inner loop to the output.}
	\label{0}
\end{figure*}

\begin{figure}[ht!]
    \centering
	\includegraphics[width=0.7\columnwidth]{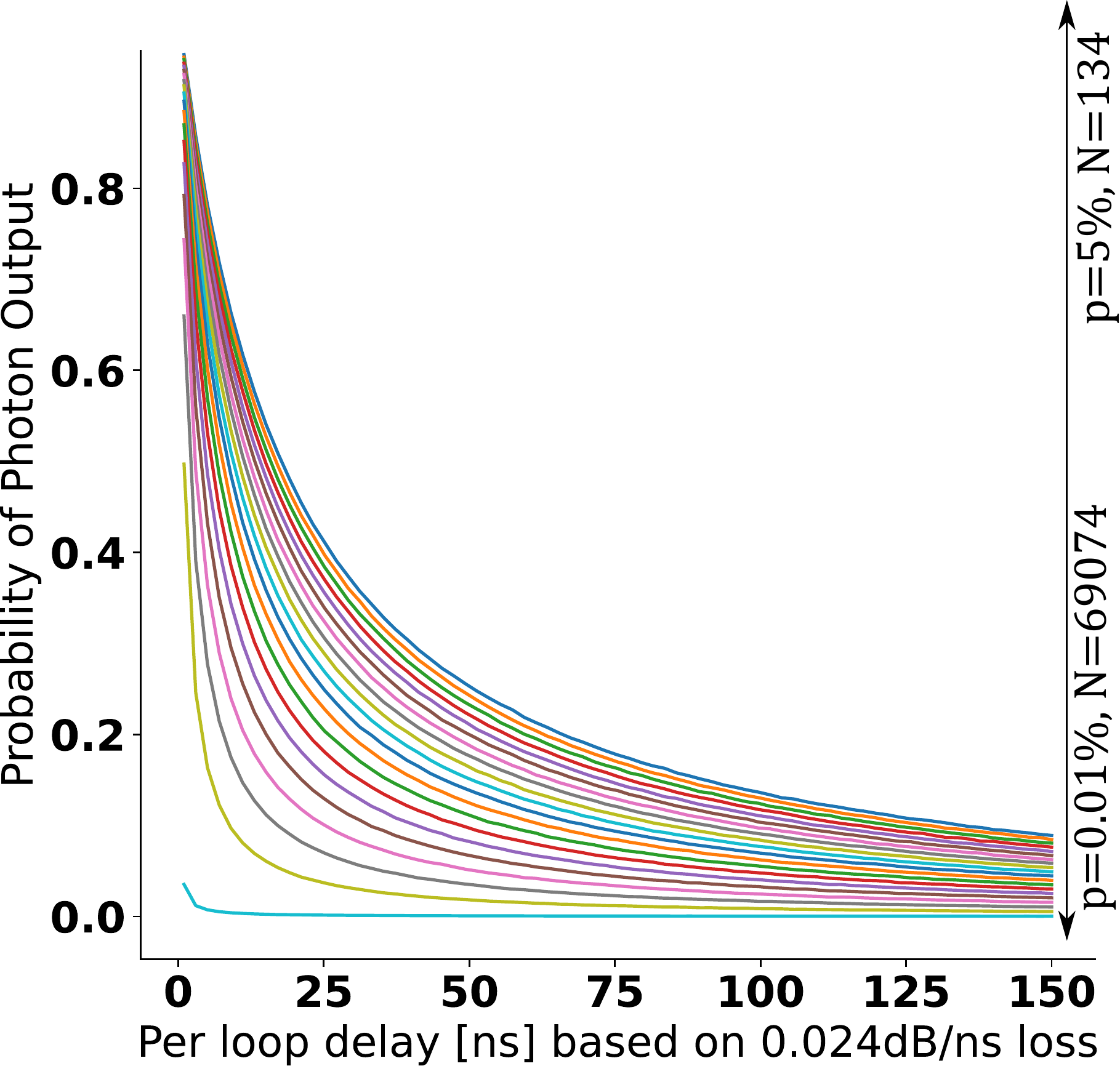}
	\caption{Photon output probability as a function of inner loop delay, number of pump cycles $N$ and photon generation probability $p$ with $p$ ranging from 0.01\% to 5\% and $N=\frac{\log{(1.0-0.999)}}{\log{(1.0-p)}}$.}
	\label{1}
\end{figure}

\begin{figure}[ht!]
	
        \includegraphics[width=1\columnwidth]{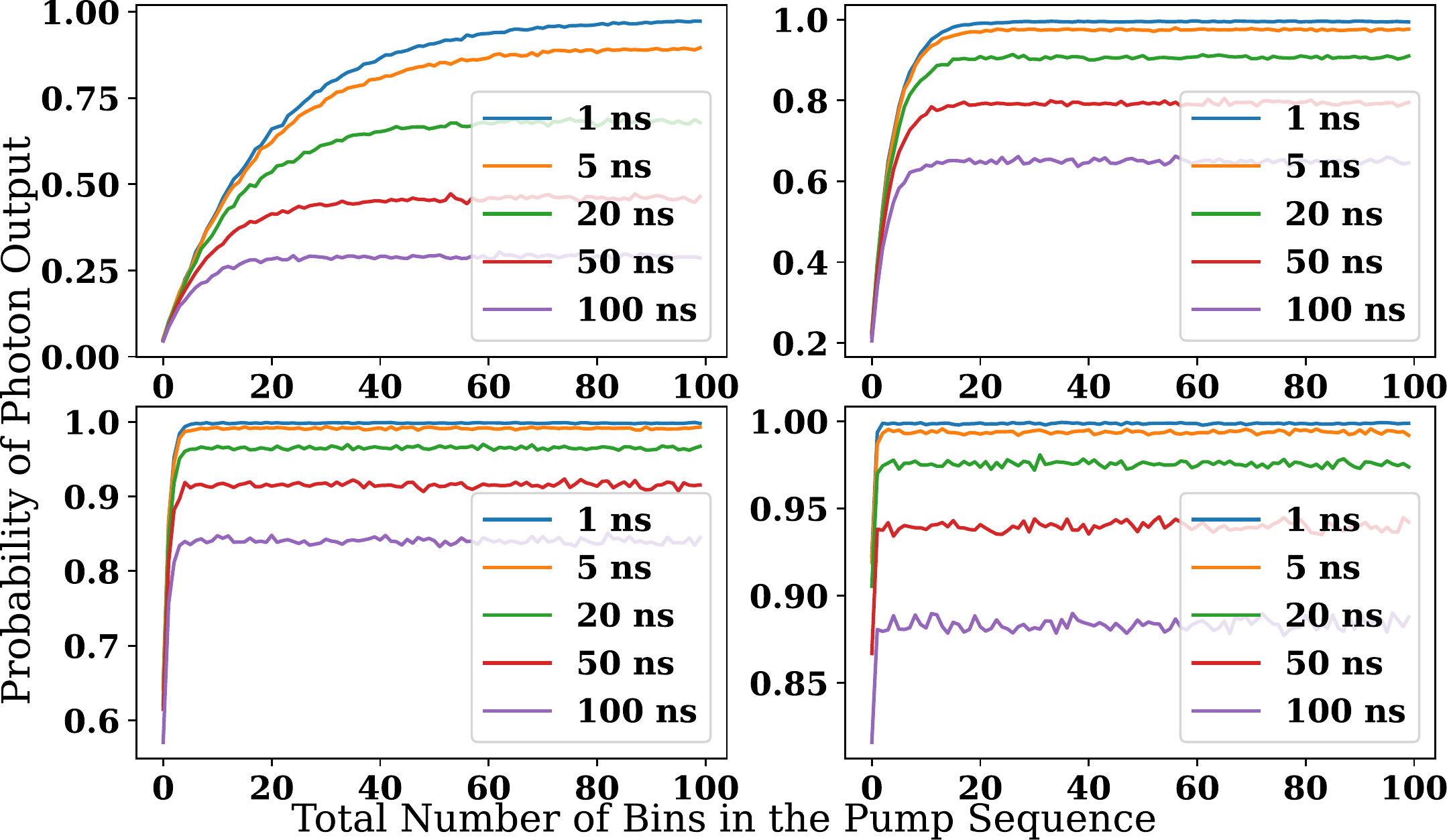}
        
	\caption{Photon output probability for $S \in \{1, 5, 20, 50\}$ photon sources and a loop delay of $T \in \{1, 5, 20, 50, 100\}$ns.} 
	\label{3}
\end{figure}

\begin{figure}[ht!]
    \centering
	\includegraphics[width=0.62\columnwidth]{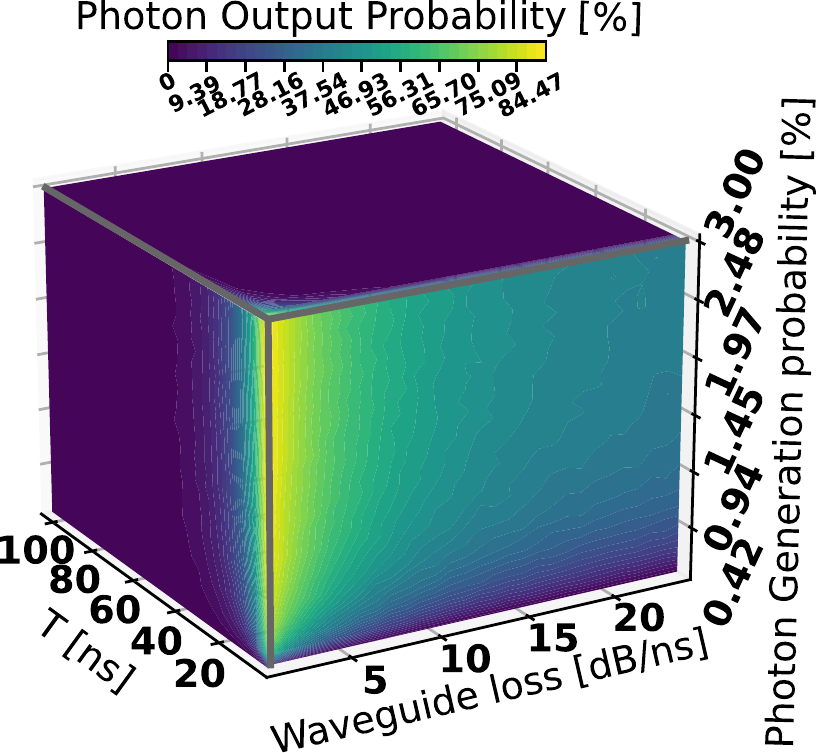}
	\caption{3D plot of several configurations of 150 SPDC sources running for 150 pump cycles showing the achievable photon output probability as a colour from dark blue to yellow for an inner loop delay ranging from 1ns to 100ns, a waveguide loss ranging from 0.001dB/ns to 24dB/ns, and a photon generation probability of a single SPDC source ranging from 0.01\% to 3\%.}
	\label{7}
\end{figure}

The proposed novel multiplexed single photon source exploits asymmetric configuration delays of switches while combining time- and space-multiplexing to improve the fidelity and excitation probability of photons.
The components of the proposed photon source are available with current technology, can be manufactured at scale and are compatible to CMOS transistors as well as telecom frequencies~\cite{10, 15}.
Therefore the photon source presents itself as an ideal candidate for the demonstration of large-scale quantum computations with linear optics.
\\
\\
The proposed photon source consists of one or more SPDC sources, which are connected to the same pump laser, single- and double-padded switches, non-photon-number-resolving detectors, delays and logic elements for processing the detector signals, e.g. CMOS transistors.
Figure~\ref{0}~shows the developed photon source with two SPDC sources that are connected via switches to an inner delay loop and the output.
Each SPDC source is connected to a detector, and to the inner loop via multiple switches that are configured at most twice. 
In each pump cycle of one SPDC source, a photon pair can be generated with probability $p$.
When the attached detector indicates that one photon pair was generated at its corresponding SPDC source, one photon in the photon pair is absorbed by the detector and the other enters the inner loop by configuring the first switch that has not been used by the proposed source, yet.
For instance, the first of the down-converted photons enters the inner loop by configuring switch $S_1$ in figure~\ref{0}, the second generated photon is entered by switch $S_2$ and so on.
The delay element directly attached to each SPDC source delays a passing photon by
\begin{equation}
    T = T_{d} + T_{c} + T_{sr},
    \label{6}
\end{equation}
where $T_d$ is the detector delay, $T_c$ is the classical processing delay and $T_{sr}$ is the previously defined switch configuration delay.
After the photon has entered into the inner loop by a switch $S_{i}$, the same switch $S_{i}$ is configured to forward newly generated photons along the outer loop to the next switch $S_{i+1}$ before the newly inserted photon traverses the inner loop once.
This is facilitated by the inner loop delay $T-\epsilon$, where $T$ is the previously defined sum of delays and $\epsilon$ is the time a photon requires to traverse the inner loop once (without being delayed by $\epsilon$). 
Thus, if a new photon is generated, it will be forwarded to the next double-padded switch that is configured to switch the inner loop photons to the outer loop and the newly generated photon into the inner loop.
Then, the same double-padded switch is configured to keep the photon in the inner loop and the next switch is used for a newly generated photon.
This is repeated until all switches of a photon source are exhausted.
The extension to multiple SPDC sources is straight forward: all detector signals are attached to classical processing, e.g. CMOS gates, to select the generated photon of one SPDC source that is forwarded to the inner loop.
After a fixed number of pump cycles, or when the component at the output of the proposed source requests, the photon from the inner loop is switched to the output.
With this setup, each switch is reconfigured at most twice, which takes full advantage of asymmetric switch configuration delays.
\\
\\
At some point during operation of the proposed single photon source, a large number of switches have been configured which warrants a reset of the photon source.
During the reset, the photon source can not output any photon and the switches are reverted to their default configuration state, which require a technology-dependent delay $T_{sf}$.
While the delay $T_{sf}$ may be large, the proposed single photon source can be designed to alleviate 
it by reducing the rate at which a reset is necessary.
The reset rate can be reduced by either connecting each SPDC source through additional switches or by using multiple single photon sources.
Both strategies extend the period after which a reset is necessary.
If the detector delay $T_d$ is larger than the switch configuration delay $T_{sr}$, detector multiplexing can be employed to reduce $T_d$ to $T_{sr}$.
\\
\\
We show the photon output probability in figure~\ref{1}~as a function of inner loop delay $T - \epsilon$, number of considered pump cycles $N$ and photon generation probability $p$ for one SPDC source attached to the outer loop.
For a decreasing photon generation probability and increasing inner loop delay, the photon output probability diminishes exponentially as detailed in figure~\ref{1}.
At a photon generation probability of $5\%$ and an inner loop delay of $9$ns the photon output probability is roughly at 66\% when performing 134 pump cycles.
This exponential decrease in photon output probability can be mitigated by introducing additional SPDC sources to the inner loop. 
\\
\\
In figure~\ref{3}, the photon output probability of the proposed single photon source is shown as a function of attached SPDC sources and pump cycles.
The number of pump cycles is shown on the x-axis and the probability of photon output is given on the y-axis.
The different lines indicate an inner loop delay ranging from $1$ns to $100$ns.
For the lowest considered inner loop delay of $1$ns, the photon output probability converges to 100\% for $5, 20$ and $50$ SPDC sources.
At one SPDC source it reaches up to 95\%.
For more than one considered SPDC sources, a sharp rise in photon output probability is observable in the region of one to ten number of pump cycles.
At higher inner loop delays, the photon loss increases and yields a diminished photon output probability.
At 5ns inner loop delay the photon loss leads to a photon output probability of less than 90\% for one source, whereas the photon output probability converges to 23\% for an inner loop delay of 100ns.
At 5, 20, and 50 SPDC sources this effect is reduced; a 100ns inner loop delay leads to a photon output probability of 62\% for 5 SPDC sources, 83\% for 20 SPDC sources, and slightly over 88\% for 50 SPDC sources.
\\
\\
Figure~\ref{7}~shows a 3D plot where the inner loop delay on the x-axis, the waveguide loss per nanosecond on the y-axis and the photon generation probability of the SPDC sources on the z-axis is varied.
The colour encodes the photon output probability of a racetrack design subject to a value assignment of the three parameters as given on their respective axes.
The number of SPDC sources and number of pump cycles is set to 150 for all data points.
It can be observed that a higher inner loop delay can mostly be compensated by a higher photon generation probability of the SPDC sources for an inner loop delay of less than 40ns. Above a inner loop delay of 40ns the photon output probability converges to a lower value regardless of the photon generation probability.
The top face of the cube shows that the waveguide loss, inner loop delay and the photon generation probability must be in a very restricted region: to allow for photon output probabilities of more than 84\%, the waveguide loss must be less than 5db/ns and the inner loop delay must be less than 10ns at a photon generation probability of 3\%.
The photon output probability converges quickly to zero for an increasing waveguide loss and inner loop delay.

\section{Error Modelling}
The inner loop delay $T$ incurs photon loss. We modeled this photon loss by the power law, i.e. 
\begin{equation}
    e^{-T\cdot I\cdot l},
\end{equation}
gives the probability that a photon is not lost while traversing the inner loop $I$ times at a waveguide loss of $l$.
The range of values investigated in the parameters of the results figures are justified by values reported in experimental setups of previous art~\cite{29, 26, 30, 28, 32, 1, 16, 3, 22}.
The work in~\cite{29, 26}~reports a waveguide loss of 0.024dB/ns, while~\cite{30}~reports a waveguide loss of 16dB/ns.
We have chosen a detector efficiency of 90\% in all of our simulations; in~\cite{30}~a detection efficiency of 78\% +- 5\% and in~\cite{28}~a detector efficiency of 95\% is reported.
Works in~\cite{32, 1}~ report a photon generation probability of 1\%, whereas the work in~\cite{30}~reports a photon generation probability of 3\%.
Works in~\cite{16, 3, 22}~report a switch configuration delay of $50$ns for the first and second configuration ($T_{sr}$) and a delay of $950$ns for the cooldown required for a third configuration. 
Photon detectors are expected to have a photon absorption delay of several picoseconds, which is negligible. However, photon detectors are also assumed to have a relatively large dead (or: reset) time that is on the order of the switch off time.
Classical processing is expected to be handled with CMOS gates, which incur a negligible delay of a few hundred of picoseconds.

\section{Auxiliary Information about the Proposed Single Photon Source}
The pump frequency is given by $T^{-1}$, where $T$ represents the interval between pumping each probabilistic source and is bounded by the response time of the heralding detector, the classical processing time and the switching time (see equation~\ref{6}). 
\\
\\
However, after a full sequence of $N$ pumping cycles is completed, the entire single photon source must be reset.  All thermal pads on all switches are deactivated and set to the default configuration.  This reset time, $T_{sf} \gg T_{sr}$ can be chosen to ensure that all switches have cleanly returned to the $\theta=0$ state without error. 
\\
\\
This consequently leads to a multiplexed source with a much lower repetition rate than the cycling time of the pumped source.
In total there are $N$ pump cycles and a reset time for each individual photon produced.
Hence the repetition rate $t_{\text{mux}}^{-1}$ of the source is given by
\begin{equation}
t_{\text{mux}} = Nt_{p} + T_r = \frac{\log(1-p_N)}{\log(1-p)}t_{p} + T_r.
\end{equation}
The repetition rate can be increased by prolonging $t_{\text{mux}}$.
This can be done by adding more switches to the inner loop that enable the routing of photons from SPDC sources to the inner loop or from the inner loop to the output.
\\
\\
Each photon source is attached to $m$ switches that can be used to switch an generated photon into the inner loop of the racetrack.
The minimum number $m$ can be computed by evaluating the inverse cumulative distribution function of the binomial distribution
\begin{equation}
    f(k, N, p) = Pr(X=k) = \binom{N}{k}p^{k}(1-p)^{N-k},
\end{equation}
where $N$ is the number of pump cycles, $p$ is specified photon generation probability and $f(k, n, p)$ is the desired probability of generating exactly $k$ photons in $N$ trials.
If all $k$ generated photons should be switched into the inner loop, the photon source requires $m := k$ switches.
We can compute the probability of generating at most $k$ photons in $N$ pump cycles, i.e. the probability that a photon source requires $m:=k$ switches by computing the cumulative distribution function of that binomial distribution
\begin{equation}
     \hat{\mathcal{F}}(k, N, p) = \sum_{i=0}^{k} \binom{N}{i}p^{i}(1-p)^{N-i}.
\end{equation}
Finally, we can determine the minimum number $m$ for a specified photon generation probability $p$ and the number of pump cycles $N$ as
\begin{equation}
    m = \hat{\mathcal{F}}^{-1}(q, N, p),
\end{equation}
where the inverse CDF $\hat{\mathcal{F}}^{-1}$ returns the number $m$ that is minimal such that
\begin{equation}
    q\leq \sum_{i=0}^{m} \binom{N}{i}p^{i}(1-p)^{N-i},
\end{equation}
with 
$q$ being the probability of being able to switch all generated photons of one SPDC source during the $N$ pump cycles into the inner loop of the racetrack.
The inverse cumulative distribution function $\hat{\mathcal{F}}^{-1}$ can be evaluated numerically~\cite{33}.

\section{Conclusion}
We have presented a multiplexed single photon source that combines space- and time-multiplexing to amplify the photon output probability.
Depending on the underlying technology parameters, the proposed single photon source can be tuned to use more time steps (pump cycles) or more SPDC sources.
The proposed single photon source can be manufactured at scale with currently available technology.
The source can exploit asymmetric configuration delays by using thermo-optic switches with two heat pads at most twice during its operation.
Simulations have shown that the proposed source can reach a photon output probability of almost 87\% at an inner loop delay of $20$ns. This work reinforces the fact that loss is by far the most dominant source of error that needs to be overcome before any scheme to improve probabilistic sources of photons can be realised.

\section{Acknowledgements}
We thank Stefanie Barz and Jeldrik Huster for helpful discussions. This work was partially funded by the Carl Zeiss foundation.
\scriptsize
\bibliographystyle{unsrt}
\bibliography{photonsource}
\end{document}